\documentclass{appolb}
\usepackage{graphicx}
\usepackage{amsmath}
\usepackage{amsfonts}
\usepackage{subcaption}
\usepackage{xcolor}
\usepackage[utf8]{inputenc} % arXiv admin
\newcommand{\be}{\begin{equation}}
\newcommand{\ee}{\end{equation}}

\begin{document}
\title{BARYON FLUCTUATIONS IN EXTENDED\\ LINEAR SIGMA MODEL%
\thanks{Presented at the Excited QCD 2020 Conference, Krynica Zdój, Poland, February 2-8, 2020.}%
}
\author{Győző Kovács\textsuperscript{1,2}, Péter Kovács\textsuperscript{1,2}
\address{\textsuperscript{1}ELTE Eötvös Loránd University,
Pázmány Péter s. 1/A,\\ H-1117, Budapest, Hungary\\ 
\textsuperscript{2}Institute for Particle and Nuclear Physics, Wigner Research Centre for Physics, H-1525 Budapest, Hungary}
\\
}
\maketitle
\begin{abstract}
The existence and the location of the critical end point (CEP) between the crossover and the first order part of the chiral phase transition in the phase diagram of the strongly interacting matter is a heavily studied area of recent particle physics. The baryon number fluctuations and related quantities such as kurtosis and other susceptibility ratios, that are assumed to be good signatures of CEP, are calculated in an (axial)vector meson extended $(2+1)$ flavor Polyakov linear sigma model (EL$\sigma$M) at zero and finite $\mu_B$. It is compared with the results of lattice as well as other effective model calculations. Divergence of the kurtosis is found at the critical end point. 
 
\end{abstract}
  
\section{Introduction}
The existence and location of a possible critical end point (CEP) between crossover and first order chiral phase transition regions on the phase diagram of strongly interacting matter is conjectured by many theoretical models \cite{Stephanov:1999zu}. If there is such a point it should be located in a region, which is unreachable neither by perturbative quantum chromodynamics (QCD) nor with lattice calculations, therefore, we only left with effective field theories. An effective, axial vector and vector meson extended $(2+1)$ flavor Polyakov linear sigma model (EL$\sigma$M) was built in \cite{Parganlija:2012fy, ref:2016}. According to symmetry considerations the Lagrangian of the model has the form:
\begin{footnotesize}
\be 
\begin{split}
\label{eq:Lag}
\mathcal{L} =&Tr \left[ \left( D_\mu \phi \right)^\dagger \left( D^\mu \phi \right) \right] - m_0 Tr \left( \phi^\dagger \phi \right) - \lambda_1 \left[ Tr \left( \phi^\dagger \phi \right) \right]^2  - \lambda_2 \left[ Tr \left( \phi^\dagger \phi \right)^2 \right] \\
&+ c_1 \left( \det \phi + \det \phi^\dagger \right) + Tr \left[ H \left( \phi + \phi^\dagger \right) \right]  - \frac{1}{4} Tr \left[ L_{\mu\nu}L^{\mu\nu}+R_{\mu\nu}R^{\mu\nu} \right]\\
& +Tr \left[\left( \frac{m_1^2 }{2} +\Delta \right) \left( L_\mu L^\mu + R_\mu R^\mu \right) \right]  +\frac{h_1}{2} Tr \left( \phi^\dagger \phi \right) Tr \left[ L_\mu L^\mu +R_\mu R^\mu \right] \\
&+ h_2 Tr \left[ \left( \phi R_\mu \right)^\dagger \left( \phi R^\mu \right) + \left( L_\mu  \phi \right)^\dagger \left( L^\mu \phi  \right) \right]  + 2 h_3 Tr \left[ R_\mu \psi^\dagger L^\mu \phi \right] \\ 
& -2 i g_2 \left(  Tr \lbrace L_{\mu\nu} \left[ L^\mu , L^\nu \right] \rbrace + Tr \lbrace R_{\mu\nu} \left[ R^\mu , R^\nu \right] \rbrace \right) + \bar{\psi} \left[ i \gamma_\mu D^\mu -\mathcal{M} \right] \psi .
\end{split}
\ee
\end{footnotesize}
Here the covariant derivatives can be written in terms of the electromagnetic fields $A^\mu$, and the gluon fields $G^\mu=g_s G^\mu_a T^a$, as
\be 
\begin{split}
D^\mu \phi &=\partial^\mu \phi - i g_1(L_\mu \phi -\phi R_\mu) -ie A^\mu \left[T_3,\phi \right], \\
D^\mu \psi &=\partial^\mu \psi - i G^\mu \psi,
\end{split}
\ee
and the field strength tensors of the left and right handed (axial) vector fields are
\be 
\begin{split}
L^{\mu\nu}&=\partial^\mu L^\nu -ie A^\mu \left[ T_3,L^\nu\right] -\lbrace \partial^\nu L^\mu -ie A^\nu \left[ T_3,L^\mu\right] \rbrace , \\
R^{\mu\nu}&=\partial^\mu R^\nu -ie A^\mu \left[ T_3,R^\nu\right] -\lbrace \partial^\nu R^\mu -ie A^\nu \left[ T_3,R^\mu\right] \rbrace,
\end{split}
\ee
The (pseudo)scalar and (axial) vector nonets are,
\begin{footnotesize}
\be \label{eq:non}
\begin{split}
&\phi =\sum_{i=0}^8 (S_i+i P_i) T_i =\frac{1}{\sqrt{2}}\begin{pmatrix}
\frac{(\sigma_N + a_0^0) +i(\eta_N +\pi^0)}{\sqrt{2}} & a_0^+ + i\pi^+ & K_0^{*+}+iK^+ \\
a_0^- + i\pi^- & \frac{(\sigma_N - a_0^0) +i(\eta_N -\pi^0)}{\sqrt{2}} & K_0^{*0}+iK^0 \\
K_0^{*-}+K^- & \bar{K}_0^{*0}+i\bar{K}^0 & \sigma_s + i \eta_s
\end{pmatrix} \\
&L^\mu =\sum_{i=0}^8 (V_i^\mu + A_i^\mu )T_i =\frac{1}{\sqrt{2}}\begin{pmatrix}
\frac{\omega_N + \rho^0}{\sqrt{2}} + \frac{f_{1N}+a_1^0}{\sqrt{2}} & \rho^+ + a_1^+ & K^{*+} + K_1^+ \\ 
\rho^- + a_1^- & \frac{\omega_N - \rho^0}{\sqrt{2}} + \frac{f_{1N}-a_1^0}{\sqrt{2}} & K^{*0} + K_1^0 \\ 
K^{*-} + K_1^- & \bar{K}^{*0} + \bar{K}_1^0 & \omega_S + f_{1S} 
\end{pmatrix}^\mu \\
&R^\mu =\sum_{i=0}^8 (V_i^\mu - A_i^\mu )T_i =\frac{1}{\sqrt{2}}\begin{pmatrix}
\frac{\omega_N + \rho^0}{\sqrt{2}} - \frac{f_{1N}+a_1^0}{\sqrt{2}} & \rho^+ - a_1^+ & K^{*+} - K_1^+ \\ 
\rho^- - a_1^- & \frac{\omega_N - \rho^0}{\sqrt{2}} - \frac{f_{1N}-a_1^0}{\sqrt{2}} & K^{*0} - K_1^0 \\ 
K^{*-} - K_1^- & \bar{K}^{*0} - \bar{K}_1^0 & \omega_S - f_{1S} 
\end{pmatrix}^\mu 
\end{split}
\ee \end{footnotesize}
The $\lbrace T_i\rbrace_{i=0}^8$ are the generators of $U(3)$, while $S_i$, $P_i$, $V_i^\mu$ and $A_i^\mu$ represents the scalar, pseudoscalar, vector and axial vector fields, respectively. Eq.~\eqref{eq:non} also shows the assignment of the physical particles except in the 0-8 sector, for which we use the so-called nonstrange-strange basis defined as,
\begin{footnotesize}
\be
\begin{split}
\varphi_N =\frac{1}{\sqrt{3}}(\sqrt{2}\varphi_0+\varphi_8), \qquad
\varphi_S =\frac{1}{\sqrt{3}}(\varphi_0-\sqrt{2}\varphi_8), \qquad \quad
\varphi \in (S_i,P_i,V_i,A_i).
\end{split}
\ee
\end{footnotesize}
In the Lagrangian Eq.~\eqref{eq:Lag} two more constant external fields $H$ and $\Delta$ appear, which have the forms
\be 
\begin{split}
H&=H_0 T_0+H_8T_8 = 
\begin{pmatrix}
\frac{h_{0N}}{2} & 0 & 0 \\
0 & \frac{h_{0N}}{2} & 0 \\
0 & 0 & \frac{h_{0S}}{\sqrt{2}} 
\end{pmatrix} \\
\Delta &=\Delta_0 T_0 +\Delta_8 T_8  = 
\begin{pmatrix}
\frac{\tilde{\delta}_N}{2} & 0 & 0 \\
0 & \frac{\tilde{\delta}_N}{2} & 0 \\
0 & 0 & \frac{\tilde{\delta}_S}{\sqrt{2}} 
\end{pmatrix} =
\begin{pmatrix}
\delta_N & 0 & 0 \\
0 & \delta_N  & 0 \\
0 & 0 & \delta_S  
\end{pmatrix},
\end{split}
\ee
The model contains 15 parameters that are fitted with $\chi^2$ method by using tree-level meson masses and decay widths. The fitting procedure and the parameters are detailed in \cite{ref:2016}.

To go to finite temperature analytic continuation to imaginary time $t\rightarrow - i\tau$ should be performed, thus the temporal part of the gluon field transformed as $G_0(t,x)\rightarrow -iG_4(\tau,x)$. Without going into the details the Polyakov loop \cite{Fukushima:2003fw, Ratti:2005jh} (path ordered Wilson loop in the temporal direction) can be defined as
\be 
L=\mathcal{P}\exp \left(i\int_0^\beta d\tau G_4 (\tau ,x)\right), \qquad L^\dagger=(L)^\dagger.
\ee
The Polyakov loop variables defined with the color traced Polyakov loops
\be \Phi(x)=\frac{1}{N_c}Tr_c L, \qquad \bar{\Phi}(x)=\frac{1}{N_c}Tr_c L^\dagger \ee
as being their thermal expectation values $\langle \Phi \rangle$ and $\langle \bar{\Phi} \rangle$,
but for simplicity we from hereafter leave the $\langle \ \rangle$ notation.

We want to study the thermodynamics of a symmetric quark matter ($\mu_u=\mu_d=\mu_s=\mu_q=1/3 \mu_B$). For this a grand potential $\Omega(T,\mu_q)$ should be obtained. 
In the mean field level approximation that is used for the evaluation of the grand potential the vacuum and thermal fluctuations of the fermions are taken into account but those for the mesons are neglected. Thus the meson potential is classical (tree-level), while the fermionic determinant is obtained after performing the functional integration over the quark field, which is evaluated for vanishing mesonic fluctuating fields. Finally the grand potential reads
\be \label{eq:GrP}
\Omega (T, \mu_q)=U (\langle M \rangle ) + U( \Phi , \bar{\Phi} ) + \Omega_{\bar{q}q}^{(0)} (T, \mu_q),
 \ee
 where $U (\langle M \rangle )$ is the tree level mesonic potential, $\Omega_{\bar{q}q}^{(0)} (T, \mu_q) $ is the fermionic contribution calculated at nonvanishing scalar-isoscalar backgrounds and vanishing mesonic fluctuations. The $U( \Phi , \bar{\Phi} )$ term is the Polyakov loop potential, which can be simply added to the grand potential, since it was treated at mean field level only so there is no integration over the gluons. One can get the field equations by minimizing the grand potential with respect to the order parameters as $\frac{\partial \Omega}{\partial \Phi}=\frac{\partial \Omega}{\partial \bar{\Phi}}=\frac{\partial \Omega}{\partial \phi_N}=\frac{\partial \Omega}{\partial \phi_S}=0$.
 
\section{Baryon number fluctuations}
The baryon number fluctuations are characterized by the higher order cumulants of the net baryon number. These can be expressed with the generalized susceptibilities , the derivatives of the pressure , which is $p=\Omega_{T=\mu_B=0} - \Omega$, with respect to the (baryo)chemical potential
\be \label{eq:chi}
\chi_n^B = \left. \frac{\partial^n p/T^4}{\partial(\mu_B /T)^n} \right|_T = \left. T^{n-4} \frac{\partial^n p}{\partial\mu_B^n} \right|_T ,
\ee
where dimensionless pressure and reduced chemical potential are used. Higher order cumulants (moments) diverges rapidly with the diverging correlation length, thus they can be good signatures of a critical endpoint \cite{Stephanov:2008qz}.

To be able to compare the experimental results to the theoretically calculated ones we should define quantities that are accessible from both sides. The ratios of the baryon number cumulants might be good candidates since the dependence on the finite volume cancels in these ratios. Note that from experimental point of view this dependence or independence is not obvious, as it is investigated in \cite{Braun-Munzinger:2016yjz}. One of these ratios is the (excess) kurtosis, that can be defined as the ratios of 4th ($k_4$) and 2nd ($k_2$) order cumulants as
\be
\kappa = \frac{k_4}{k_2^2} .
\ee  
Since
\be \begin{split}
k_2&=VT\frac{\partial^2 p}{\partial \mu_B^2} =VT\frac{T^4}{T^2} \chi_2^B = VT^3  \chi_2^B, \\
k_4&=VT^3\frac{\partial^4 p}{\partial \mu_B^4} =VT^3\frac{T^4}{T^4} \chi_4^B = VT^3  \chi_4^B, 
\end{split} \ee
and $k_2 = \sigma^2$ is the variance, it can been rewritten as
\be
 \kappa\sigma^2 = \sigma^2 \frac{k_4}{k_2^2}= \frac{\chi_4^B}{\chi_2^B }
\ee
This quantity, $\kappa\sigma^2$ that we actually call kurtosis. We calculated the susceptibilities numerically with finite difference method. 

\section{Results}

\begin{figure}[htp]
        \centering
        \begin{subfigure}[b]{0.475\textwidth}
            \centering
            \includegraphics[width=\textwidth]{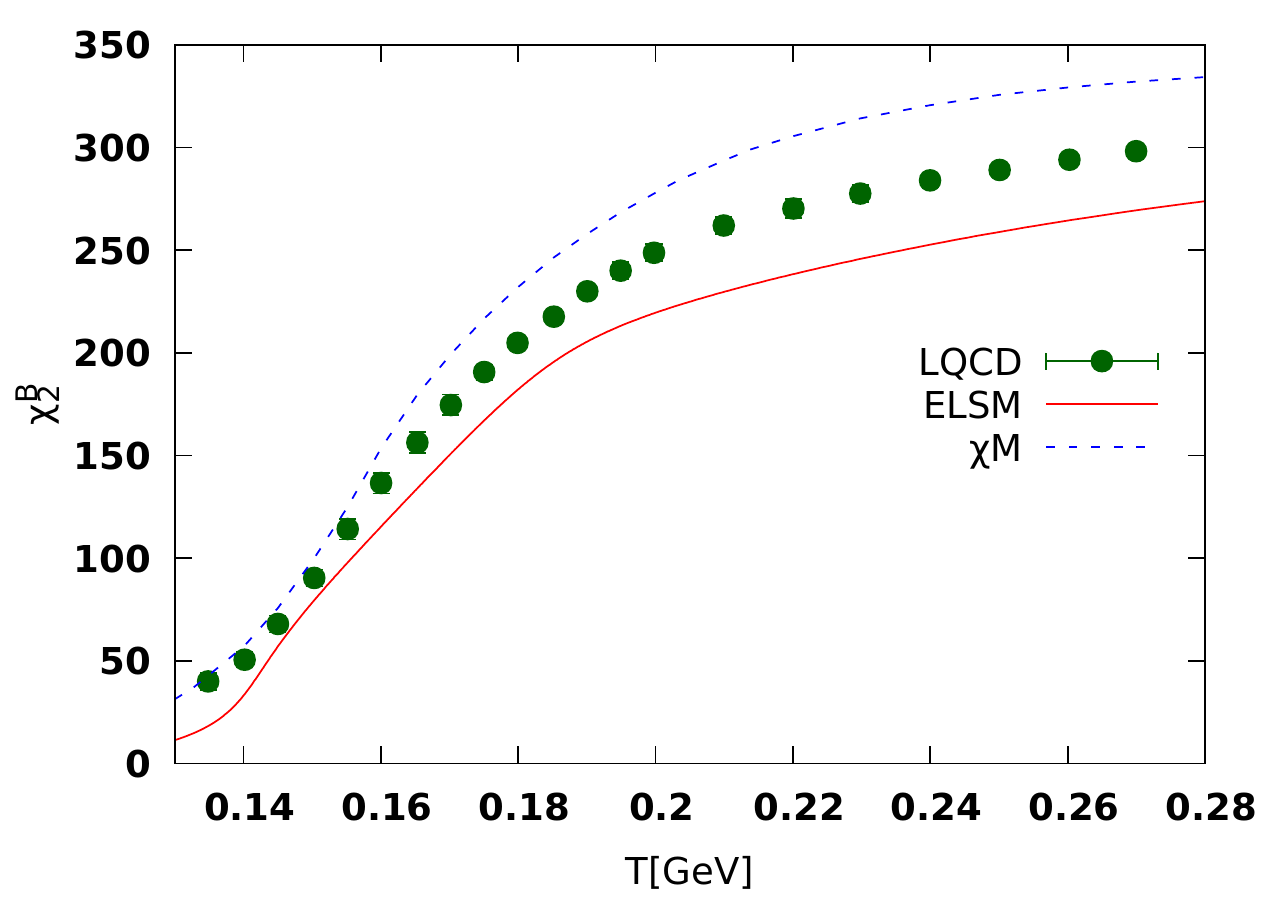}
        \end{subfigure}
        \hfill
        \begin{subfigure}[b]{0.475\textwidth}  
            \centering 
            \includegraphics[width=\textwidth]{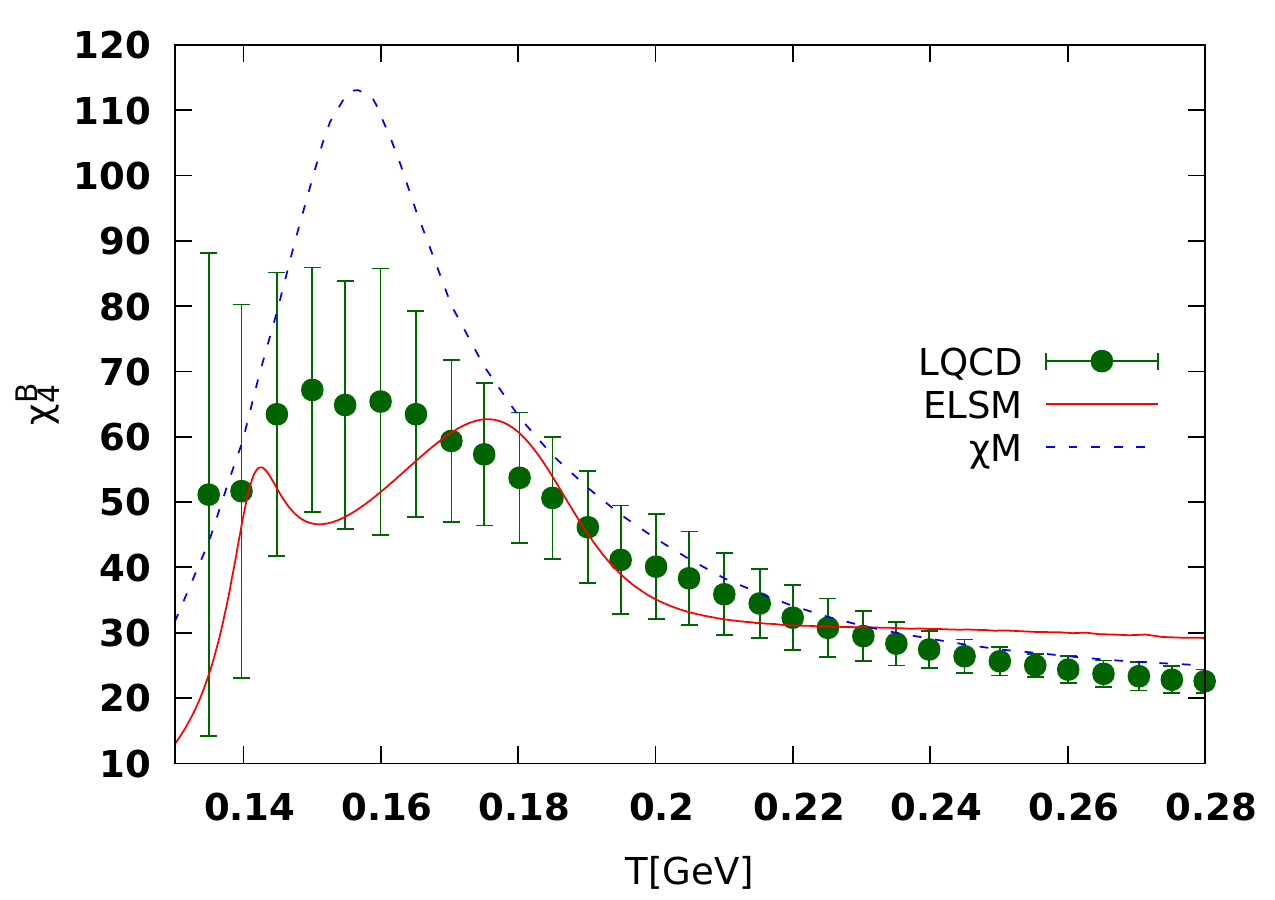}
        \end{subfigure}
        \vskip\baselineskip \vspace*{-0.5cm}
        \begin{subfigure}[b]{0.475\textwidth}   
            \centering 
            \includegraphics[width=\textwidth]{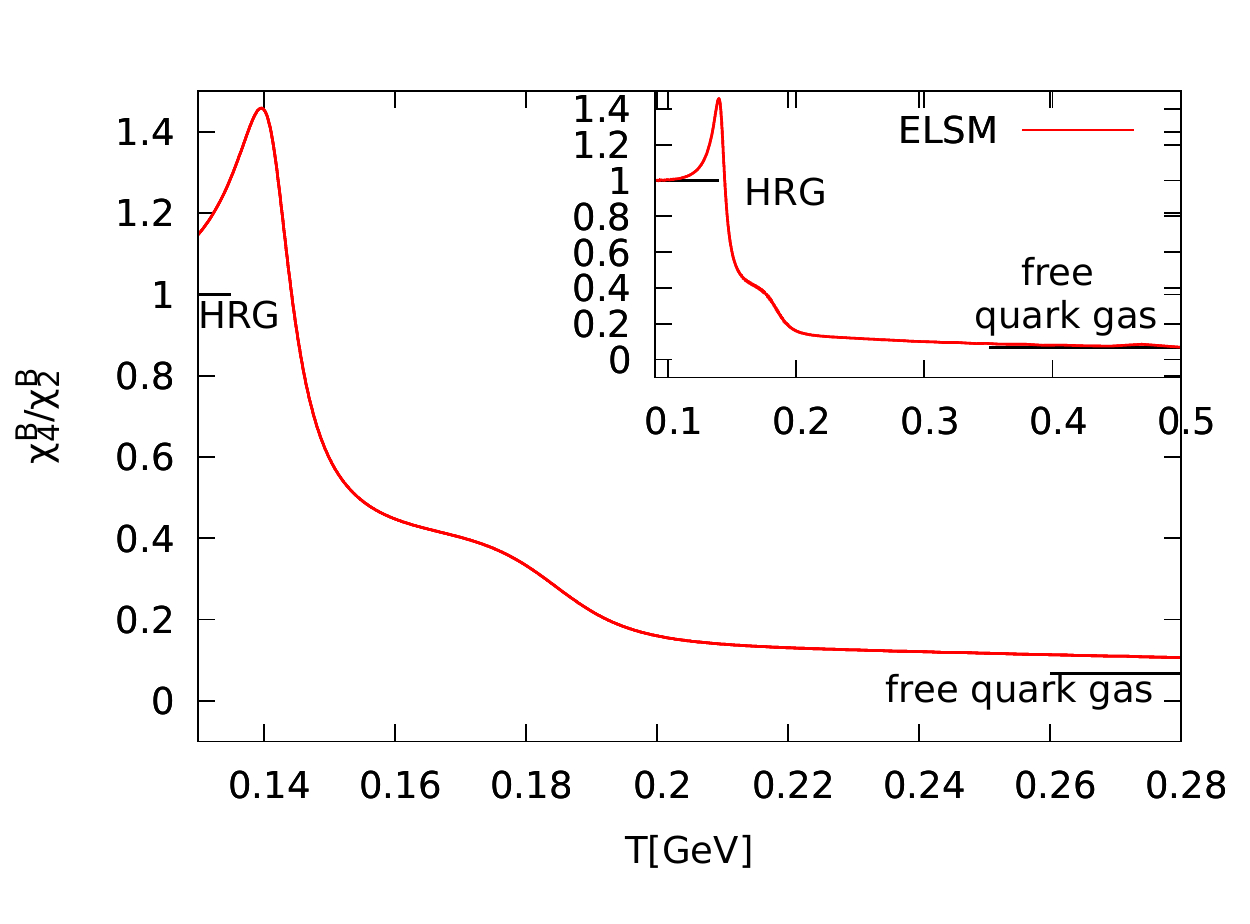}
        \end{subfigure}
        \quad
        \begin{subfigure}[b]{0.475\textwidth}   
            \centering 
            \includegraphics[width=\textwidth]{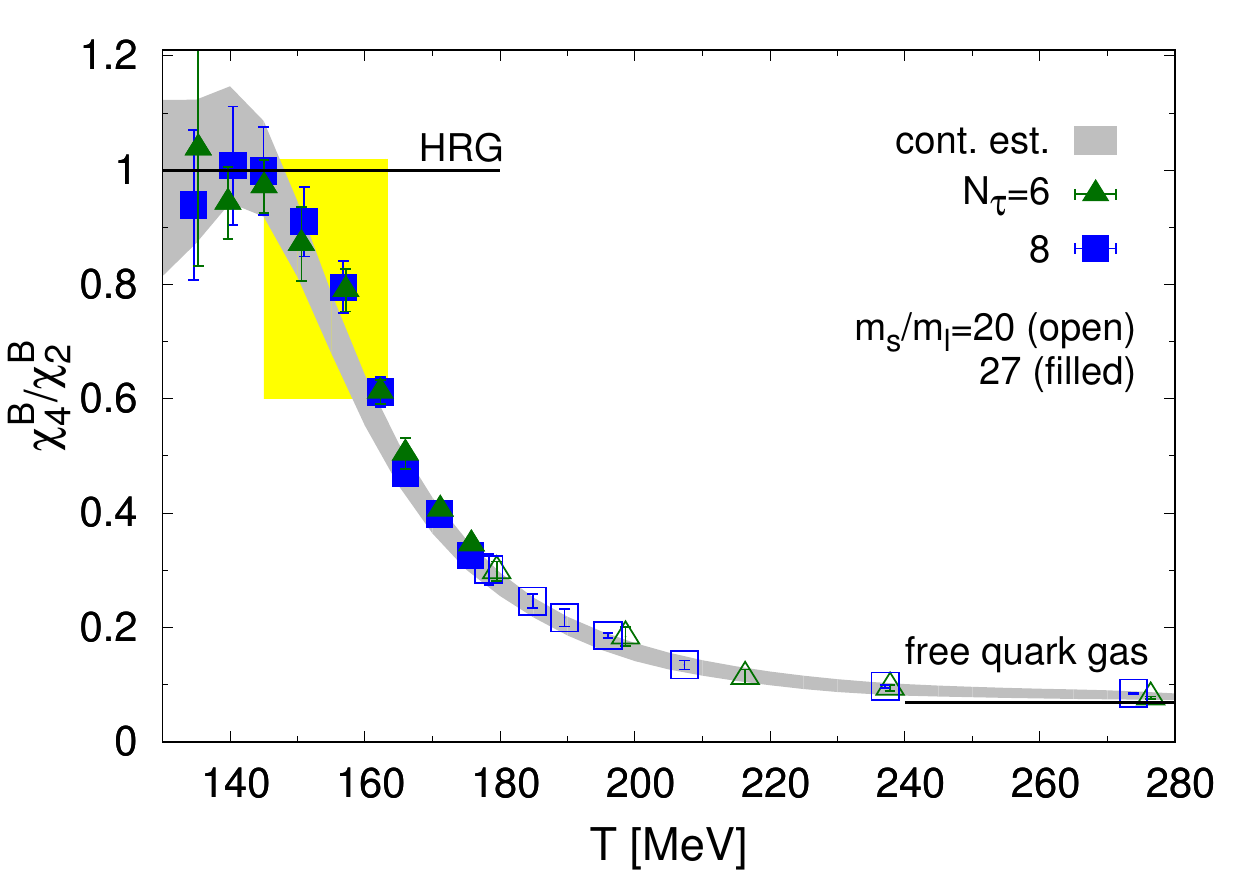}
        \end{subfigure}
        \caption{ {\footnotesize $\chi_2^B$ as function of temperature in the EL$\sigma$M, in the chiral matrix model, and on the lattice \cite{ref:PisSko} (top left). Temperature dependence of $\chi_4^B$ compared again to the chiral matrix model and lattice data (top right). The kurtosis in the ELSM (bottom left) and on the lattice \cite{Bazavov:2017dus} (bottom right). The inset of the bottom left figure contains the same curve depicted on a larger scale to show how the HRG and free quark gas limits are approached.}}
        \label{fig:R}
\end{figure}

Now we present our results in the EL$\sigma$M both at zero and finite $\mu_B$. For zero chemical potential the 2nd and 4th order susceptibilities and the kurtosis are shown in Figure \ref{fig:R}.
    
%Figure

As it can be seen our results are compatible with the lattice results. It is worth to note that in the case of the EL$\sigma$M there are a double peak structure around the phase transition (see {\it e.g.} top right of Fig. \ref{fig:R}). The reason behind this behavior is that we have four order parameters. These, namely the Polyakov loop parameters $\Phi,\ \bar{\Phi}$ and the scalar-isoscalar vacuum expectation values $\phi_N,\ \phi_S$ changes at slightly different temperatures, which cause the separation of peaks for the different order parameters. 

\begin{figure}[htp]
  \centering
  \includegraphics[width=\textwidth]{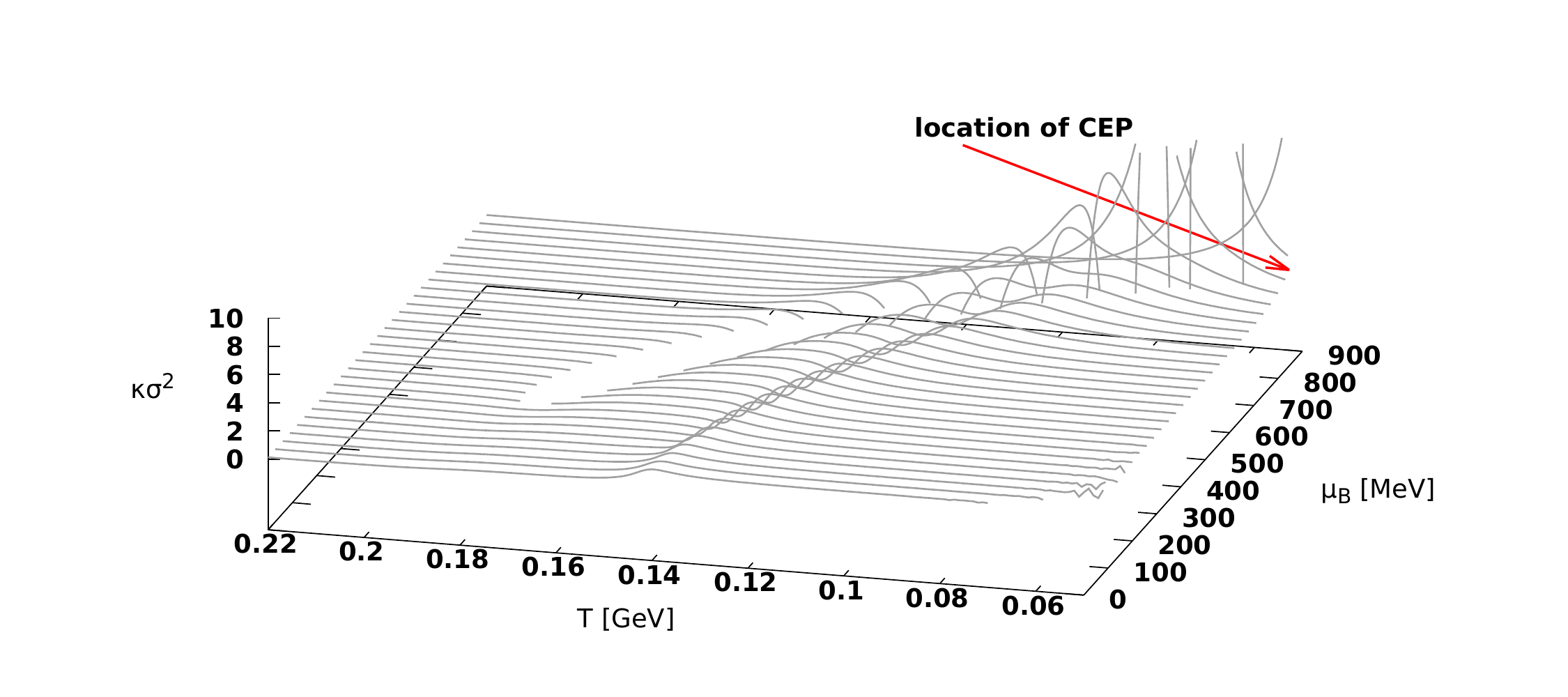}
  \caption{ {\footnotesize The 3D plot of the kurtosis as a function $T$ and $\mu_B$. The arrow points toward our prediction of the CEP at $\mu_B=0.885\ GeV$ and $T=0.052\ GeV$.}  }  
  \label{fig:T}
\end{figure}

It is also worth to investigate the behavior of kurtosis at finite $\mu_B$, which is shown on Fig. \ref{fig:T}. 
Going towards the predicted critical end point (CEP) the value of the kurtosis at the phase transition increases significantly and -- as it is expected -- even diverges at the CEP.
	
%\section{Summary and outlook}
%We calculated the fluctuations of net baryon number in ELSM in zero and finite $\mu_B$. We pla

\section*{Acknowledgement}

Gy. Kovács acknowledges support by the NRDI fund of Hungary, financed under the FK\textunderscore19 funding scheme, project no. FK 131982, while his conference participation was also supported by the framework of COST Action CA15213 THOR. P. Kovács acknowledges support by the János Bolyai Research Scholarship of the Hungarian Academy of Sciences and was also supported by the ÚNKP-19-1 New National Excellence Program of the Ministry for Innovation and Technology.

\bibliography{sample}

\end{document}